\documentclass{article}

\usepackage{arxiv}
\usepackage[utf8]{inputenc}
\usepackage[T1]{fontenc}
\usepackage{amsmath,amssymb,amsthm,booktabs,multirow,bm}
\usepackage{graphicx}
\usepackage[numbers,sort&compress]{natbib}
\usepackage{hyperref}

\newtheorem{definition}{Definition}

\renewcommand{\headeright}{Preprint}
\renewcommand{\undertitle}{}
\renewcommand{\shorttitle}{How local rules generate emergent structure in cellular automata}

\begin{document}

\title{How local rules generate emergent structure in cellular automata}

\author{
  Manuel Pita\thanks{ORCID: \href{https://orcid.org/0000-0003-2180-6823}{0000-0003-2180-6823}} \\
  Artificial Intelligence, Social Interaction and Complexity Laboratory, \\CICANT, Universidade Lus\'{o}fona\\
  Campo Grande 376, Lisbon, Portugal\\
  \texttt{manuel.pita@ulusofona.pt}
}

\date{March 31, 2026}

\maketitle

\begin{abstract}
Cellular automata generate spatially extended, temporally persistent emergent structures from local update rules. No general method derives the mechanisms of that generation from the rule itself; existing tools reconstruct structure from observed dynamics. This paper shows that the look-up table contains a readable causal architecture and introduces a forward model to extract it. The key observation in elementary cellular automata (ECA) is that adjacent cells share input positions, so the prime implicants of neighbouring transitions overlap. That overlap can couple the transitions causally or leave them independent. We formalize each pairwise interaction as a \emph{tile}. A finite-state, \emph{tiling} transducer, $\mathcal{T}$, composes tiles across the CA lattice, tracking how coupling and independence propagate from one cell pair to the next. Structural properties of $\mathcal{T}$ are used to classify ECA rules that can sustain regions of causal independence across space and time. We find that, in the 88 ECA equivalence classes, the number of local configurations at which coupling is structurally impossible---computable from the look-up table---predicts the prevalence of dynamically decoupled regions with Spearman $\rho = 0.89$ ($p < 10^{-31}$). The look-up table encodes not just what a rule computes but where it distributes causal coupling across the lattice; the framework reads that distribution forward, from local logical redundancy to emergent mesoscopic organization.
\end{abstract}


\section{Introduction}

Some of the most complex collective behaviours observed in nature arise from remarkably simple local interactions. Flocking, pattern formation, synchronization and turbulence all emerge from rules that operate on immediate neighbours alone. Cellular automata (CAs) distil this phenomenon to its essence. A single Boolean function, applied identically at every cell in a regular lattice, generates spatiotemporal structures---domains, particles, gliders, complex interactions---that span the lattice and persist across time. The emergence of such structures must be traceable to the rule's look-up table (LUT). Yet after four decades of study, no general method exists for reading the mechanisms that produce them directly from the table's entries.

The difficulty is not a lack of tools but a shared direction of inference. The epsilon-machine~\cite{crutchfield1989,hanson1992} reconstructs a finite-state transducer from observed spacetime data that recognizes regular domains. Shalizi et al.'s local statistical complexity filter~\cite{shalizi2006} measures the predictive information at each spacetime cell. Lizier's information-theoretic decomposition~\cite{lizier2012} partitions the dynamics into local storage, transfer and modification components. These tools are powerful and have produced critical insights that have advanced the theory of complex systems. But they all work \emph{backward}: from observed dynamics to inferred features of the emergent patterns---revealing \emph{what} structures a rule produces. The unanswered question remains: \emph{how} does the rule's Boolean logic produce them?

Langton's $\lambda$ parameter~\cite{langton1990} approaches the question from the rule side, locating the region of rule space where complex behaviour concentrates. Langton also proposed that computation in CAs requires three primitive functions: storage, transmission and modification of information. Langton's $\lambda$ parameter is a statistical summary of the look-up table. It identifies \emph{where} in rule space to look, not \emph{how} a given rule supports those functions. 

This paper shows that the rule table itself contains a readable causal architecture, and introduces a forward model to extract it. The starting point is wildcard schema redescription~\cite{marquespita2011,marquespita2013}---the decomposition of a Boolean function into its prime implicants via Quine--McCluskey minimization~\cite{quine1952,mccluskey1956}. Prime implicants are the maximally compressed input patterns that determine each output; their wildcard positions mark causally irrelevant inputs. Schema redescription reveals the causal structure of individual transitions in Boolean automata. The framework developed here extends that characterization to the CA spatiotemporal dynamics. The key observation is that, in elementary CAs, adjacent cells share exactly two input cells---so the prime implicants of neighbouring transitions overlap, and the overlap can either couple the two transitions causally or leave them independent. We define an ordered pair of adjacent prime implicants whose overlap positions are not contradicted as a \emph{tile}; whether the shared positions are required inputs or irrelevant to each transition determines whether the tile couples or decouples the adjacent cells. The set of all tiles---the \emph{tile catalogue}---is the rule's complete repertoire of pairwise coupling interactions, computable from the look-up table alone. The lattice supplies adjacency; the prime-implicant structure determines where that adjacency becomes direct causal coupling, where it is optional, and where it is structurally absent.

Tiles do not compose freely: the overlap demands left unresolved by one tile constrain which tiles may follow. A finite-state transducer we call the \emph{tiling transducer} $\mathcal{T}$ tracks these constraints across the lattice. At each interface between adjacent cells, the framework determines whether coupling is structurally guaranteed, structurally impossible, or dependent on which prime implicants are chosen---a three-way classification computable from the look-up table. Two structural properties of $\mathcal{T}$ then identify which rules can sustain regions of structural independence that persist across both space and time, without needing to examine the dynamics.

The difference between forward and backward inference is concrete. Elementary cellular automata (ECA) $\phi_{18}$ and $\phi_{30}$ both produce spatiotemporal patterns that appear random to the unaided eye~\cite{wolfram1983,wolfram2002}. Their spacetime diagrams are superficially similar: irregular, aperiodic, resistant to visual parsing. We show that their causal architectures differ fundamentally. Computational mechanics~\cite{hanson1997} identifies the apparent randomness of $\phi_{18}$ as a regular domain threaded by localized particles---a finite-state description reconstructed from the dynamics. $\phi_{30}$ has resisted all such decomposition~\cite{wolfram2002,wolfram2019}. The backward tools describe this difference. The tiling transducer explains it: the distribution of causal redundancy across the prime implicants of $\phi_{18}$ produces a repeating sequence of independent cells that self-sustains across the lattice---the mechanism behind the regular domain. The prime implicants of $\phi_{30}$ lack this structural property.

Across all 256 ECAs (88 equivalence classes under reflection and input--output complement), the number of local configurations at which coupling is structurally impossible---computable from the look-up table alone---predicts the prevalence of spatially extended, temporally stable regions of causal independence in the dynamics, with Spearman $\rho = 0.89$ ($p < 10^{-31}$; 50-seed mean). The framework links microscopic logical structure---the distribution of causal redundancy across the rule's prime implicants---to emergent mesoscopic organization.

The argument proceeds as follows. \S\ref{sec:background} reviews existing tools for collective information processing in cellular automata, identifying the look-up-table-to-structure gap. \S\ref{sec:schema-tiling-framework} develops the framework: tiles, $\mathcal{T}$, the three-way bond classification, wildcard-emission cycles and self-consistency. \S\ref{sec:case-studies} demonstrates the framework on $\phi_{18}$ and $\phi_{30}$. \S\ref{sec:hierarchy} analyses all 256 ECA rules (88 equivalence classes). \S\ref{sec:dynamical-validation} validates the structural predictions dynamically. \S\ref{sec:discussion} discusses the framework's diagnostic power and its limitations.


\section{Background}
\label{sec:background}

\subsection{Cellular automata and emergent behaviour}

A cellular automaton (CA) is a discrete dynamical system defined on a regular lattice of identical cells, where $s_{i}$ denotes the state of cell $i$. The lattice is updated synchronously: a deterministic look-up table (LUT) maps each cell's fixed local neighbourhood at time $t$ to its next state at $t+1$. Elementary cellular automata (ECA) are the one-dimensional, binary-state case with periodic boundary conditions and neighbourhood radius $r=1$: cell $s_i$ updates from the triple $(s_{i-1}, s_i, s_{i+1})$.
The concrete state of cell $s_i$'s local neighbourhood is the triple $\eta_i = (s_{i-1}, s_i, s_{i+1}) \in \{0,1\}^3$.
The LUT (CA rule), denoted by $\phi$, maps each of the $2^3 = 8$ possible neighbourhood configurations $\eta_i$ to a binary output.
Since the LUT maps the next state of eight input neighbourhoods, there are 256 possible ECAs.
We refer to the ECAs using Wolfram's numbering convention, which reads the eight output bits as a binary number \cite{wolfram1983}.
Under left--right reflection and input--output complement (swapping $0 \leftrightarrow 1$ in both inputs and outputs), the 256 ECAs reduce to 88 essential equivalence classes \cite{li1990}.

Wolfram~\cite{wolfram1983} classified ECA behaviour into four classes: Class~I ECAs, whose dynamics converge to homogeneous configurations, Class~II, to separated simple stable or periodic structures, Class~III, to chaotic patterns, and Class~IV, to complex localized structures that can be long-lived. The classification describes the phenomenology of ECA dynamics but does not explain the mechanism by which a given rule produces its class of behaviour.

\begin{figure}[!ht]
\centering
\includegraphics[width=0.85\columnwidth]{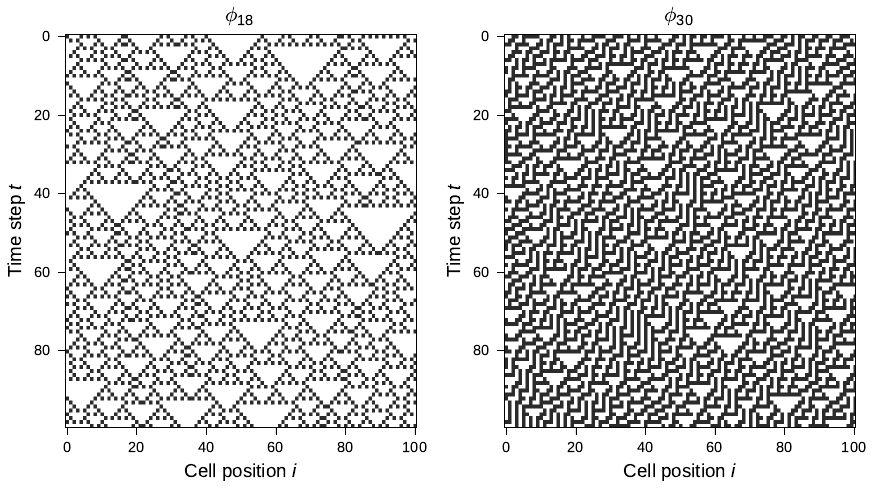}
\caption{\label{fig:spacetime_intro} Spacetime diagrams of $\phi_{18}$ and $\phi_{30}$ computed from the same random initial configuration ($N = 101$ cells, $T = 200$ steps. The first 50 transient steps were discarded. Black (white) cells are in state~1 (0). Both rules produce aperiodic, apparently random patterns from identical initial data.}
\end{figure}

This paper uses two ECAs as running examples, $\phi_{18}$ and $\phi_{30}$. Both belong to Wolfram's Class~III, rules whose spacetime diagrams appear aperiodic and random from arbitrary initial configurations~\cite{wolfram1983,wolfram2002}. Figure~\ref{fig:spacetime_intro} shows their spacetime dynamics from the same random initial configuration. The diagrams are superficially similar: irregular, aperiodic, resistant to visual parsing.

\subsection{Collective information processing}
\label{sec:collective-information-processing}

Langton's $\lambda$ parameter~\cite{langton1990} identifies the region of rule space where complex behaviour concentrates---near a phase transition between ordered and chaotic dynamics. Langton proposed that computation requires three primitive functions: storage, transmission and modification of information. The parameter locates \emph{where} in rule space to look but does not explain \emph{why} a given rule supports these functions. Several approaches have moved beyond phenomenological classification by reconstructing CA structure directly from observed spacetime dynamics.

Crutchfield and Young's~\cite{crutchfield1989} epsilon-machine infers a minimal finite-state machine from a CA's spacetime configurations~\cite{hanson1992}. The machine identifies \emph{regular domains}: spatial patterns whose structure repeats periodically across the lattice and persists from one time step to the next. The localized boundaries between domains are \emph{particles} that propagate through the lattice. Where particles collide, they annihilate or produce new particles~\cite{hordijk1998}. The epsilon-machine thus decomposes the spacetime field into three structural elements (domains, particles and interactions), each defined by its role in the spatiotemporal pattern.

Shalizi et al.~\cite{shalizi2006} introduced a local statistical complexity filter that operates pointwise, measuring at each spacetime cell the information needed to predict its future from its past lightcone. Domains, being repetitive, require little predictive information and appear as low-complexity regions; particles and collisions require more and appear as high-complexity boundaries.

Lizier, Prokopenko and Zomaya~\cite{lizier2012} decomposed each cell's dynamics into three information-theoretic quantities. The decomposition measures how much of a cell's next state is predictable from its own past (active information storage), how much is contributed by its neighbours (transfer entropy), and how much arises only when both sources combine (information modification).

More recently, Rupe and Crutchfield~\cite{rupe2018} moved from global grammars of spatial configurations to pointwise labelling of individual spacetime cells. Each cell is assigned a local causal state, the equivalence class of all past lightcones that yield the same predictive distribution over futures. A domain is a spacetime region in which the same local causal state repeats from cell to cell; a coherent structure is a localized, persistent region where the labelling changes. Because the equivalence classes are built entirely from observed lightcones, the approach requires no knowledge of the LUT.

These approaches are complementary, but they share a direction of inference: all reconstruct structure from observed dynamics. None derives structure from the look-up table. A backward reconstruction describes \emph{what} structures appear; only a forward derivation from the LUT can explain \emph{how} they emerge.

Applied to the running examples, these frameworks reveal that the superficial similarity between $\phi_{18}$ and $\phi_{30}$ conceals a fundamental architectural difference. For $\phi_{18}$, the epsilon-machine identifies a single regular domain: the $(0\Sigma)^*$ pattern, in which every other cell is zero, and the intervening cells can take either value~\cite{hanson1997}. Particles appear as localized boundaries between shifted copies of this domain, executing random walks whose starting positions are set by the non-zero cells in the initial condition. Lizier et al.'s decomposition confirms the picture~\cite{lizier2012}. Domain cells are largely predictable from their own past; particle cells are driven by their neighbours. The apparent randomness of $\phi_{18}$ decomposes into identifiable structural elements. For $\phi_{30}$, no comparable decomposition has been achieved. No regular domain has been identified in its spacetime dynamics, and without domains, particles cannot be defined. Basic structural properties of $\phi_{30}$ remain open problems~\cite{wolfram2019}.

\subsection{Prime implicants, wildcards and enputs}
\label{sec:prime-implicants}

A Boolean look-up table enumerates input–output pairs but does not reveal which inputs each transition requires. Boolean minimization makes this structure explicit. Consider, for example, a three-input OR function. If any input is 1, the output is 1; the remaining two inputs are thus irrelevant. The Quine--McCluskey algorithm~\cite{quine1952,mccluskey1956} applies this reasoning to any Boolean function, finding its complete set of prime implicants (PIs)---the set of minimal combinations of input values that make the other inputs irrelevant.

A prime implicant is a wildcard generalization of an input pattern that cannot be made more general and still guarantee the output value: specifying fewer inputs would break the guarantee. Each PI is like a LUT entry, but the inputs are defined over $\{0, 1, \#\}$, where $\#$ (the wildcard) marks an input that becomes irrelevant to the output. We refer to the non-wildcard positions as the \emph{essential inputs}, or \emph{enputs}---the values of which are minimally necessary to determine the output. Wildcards and enputs are properties of the rule, not artefacts of the minimization~\cite{marquespita2011}. In CAs, PIs reveal which inputs control the state transitions.
We denote the set of PIs as $F' = \{f'_0, \ldots, f'_m\}$, where $m$ is the number of prime implicants. The set is ordered by ascending wildcard count, then lexicographically. We use $|f'_k|_\#$ to refer to the wildcard count of the $k$th prime implicant.

\subsection{Canalization and causal structure in automata networks}
\label{sec:canalization}

Waddington~\cite{waddington1942} introduced canalization in biology to explain why phenotypic traits remain stable despite genetic and environmental variation. The PI decomposition (\S\ref{sec:prime-implicants}) makes canalization explicit in Boolean functions.
A PI with $|f'_k|_\#$ wildcards means that many of the inputs described by the PI are irrelevant; the output is canalized for those input combinations. In the three-input OR, a single input in state~1 renders the other two irrelevant: the output is 2-canalized for the four input combinations that PI covers.

In Boolean networks (BNs), used as models of genetic regulation~\cite{kauffman1993,kauffman2004}, Marques-Pita and Rocha~\cite{marquespita2013} showed that enputs identify which minimal subsets of nodes, in specific state combinations, exert contextual \emph{control} over the dynamics. Their framework constructs a \emph{dynamics canalization map} (DCM): a graph of enput-driven transitions that enables forward inference from incomplete initial configurations. The DCM identifies minimal control conditions and dynamical modules. They also introduced \emph{effective connectivity}, a per-node measure of the expected number of inputs needed to determine the node's output after redundancy is removed.
Gates et al.~\cite{gates2021} used effective connectivity to produce an \emph{effective graph} that reweights each edge in a BN by the causal influence the upstream node actually exerts on the downstream node's transitions. The effective graph reveals control pathways invisible in the structural wiring. Manicka, Marques-Pita and Rocha~\cite{manicka2022} then showed that effective connectivity predicts whether a network operates in the critical regime between order and chaos.
In previous work, Marques-Pita and Rocha~\cite{marquespita2008, marquespita2011} applied canalization to cellular automata. They evolved CA rules to perform the majority classification task~\cite{mitchell1993} with high performance by constraining the prime-implicant structure of the evolved rules to have a `process symmetry' property.

These contributions characterize the causal structure of individual nodes and pairwise connections in BNs with arbitrary topology, or improve the evolution of CAs that perform collective computation. The DCM could, in principle, support forward derivation of causal architecture. However, Marques-Pita and Rocha applied it to trace control from specific initial states to attractor configurations, not to derive the causal structure directly from the BN structure and update functions. The heterogeneity of BNs (each node with a distinct function and in-degree) required that such analysis be computed node by node and network by network. 

A cellular automaton is uniform: every cell applies the same update function to a fixed local neighbourhood (radius $r=1$ in ECAs). The prime implicant decomposition is performed once, for the rule, and spatial composition follows from the lattice alone. The uniformity and single Boolean function of a CA create the conditions for a forward derivation of causal architecture from the rule, and this paper constructs the machinery to carry it out.

The causal claim in this framework adopts an interventionist reading of the Boolean function~\cite{woodward2003,pearl2000}. The PI decomposition provides the counterfactual structure interventionist causation requires: changing an enput changes the output; changing a wildcard leaves it invariant. In a deterministic system, the Boolean function is the complete mechanism, and this counterfactual dependence constitutes causation. This sense of `causal' differs from the causal states of computational mechanics~\cite{rupe2018,crutchfield1989}, which are equivalence classes of pasts that are predictively sufficient for the future. That criterion is information-theoretic; the criterion here is interventionist. The two are complementary: one reads causal structure from the dynamics, the other from the rule.


\section{The CA tiling framework}
\label{sec:schema-tiling-framework}

Prime implicants (\S\ref{sec:prime-implicants}) reveal what inputs, in which value combinations, are minimally necessary to determine the output of a Boolean function.
This allows us to frame an interventionist question within CA dynamics: if the state of a cell $s_{i}$ were unknown, could the transitions that depend on it still be resolved?
If the PIs for transitions that depend on $s_{i}$ all have a wildcard for the corresponding input, the answer is yes.
The relevant state transitions do not depend on that cell's state.

Every pair of adjacent cells in an ECA shares exactly two inputs. Whether a shared input $s_i$ is an enput to neither, one, or both transitions determines the causal relationship between the two cells:

\begin{enumerate}
\item Neither: both cells determine their next state without $s_i$.
\item One only: both cells update independently.
\item Both: neither can be resolved independently. The two transitions are \emph{causally coupled} through that shared enput.
\end{enumerate}

Prime implicants thus characterize the causal coupling architecture of a CA lattice at any given configuration. The complete repertoire of possible causal architectures is derivable from the LUT alone. This section develops the formal apparatus to extract it.
We introduce the concept of \emph{tile} to formalize pairwise PI coupling between adjacent transitions (\S\ref{sec:tiles}), a finite-state transducer that composes tiles across the lattice (\S\ref{sec:tiling-transducer}), and a scalar measure of the rule's global decoupling (\S\ref{sec:three-way}).


\subsection{Tiles, overlap resolution and coupling weight}
\label{sec:tiles}

\begin{figure}[!ht]
\centering
\includegraphics[width=0.7\columnwidth]{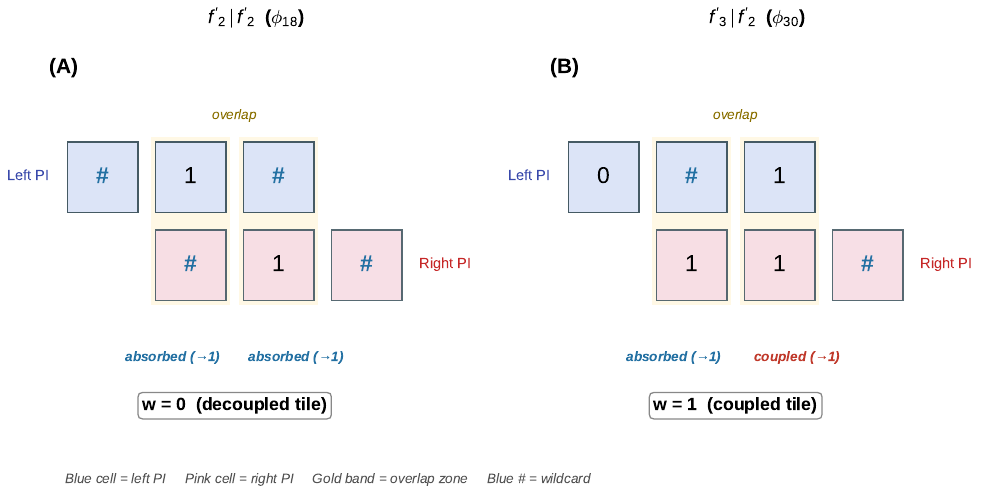}
\caption{\label{fig:overlap} Overlap resolution for two example tiles (list of PIs in Table \ref{tab:PIs}). (A)~$f'_2 \mid f'_2$ in $\phi_{18}$: both overlaps resolve by absorption ($w = 0$, decoupled tile); (B)~$f'_3 \mid f'_2$ in $\phi_{30}$: one overlap absorbed, one coupled ($w = 1$, coupled tile). Blue (pink) cells represent left (right) PI; the overlap cells, $\mathcal{C}$, are highlighted in yellow.}
\end{figure}

In ECAs the next state of a cell $s_i$ depends on the set of input cells $N_i = \{s_{i-1}, s_{i}, s_{i+1}\}$. $N_i$ denotes the set of input cells; $\eta_i$ (\S\ref{sec:background}) denotes their concrete state assignment.
The neighbourhoods of two adjacent cells overlap on exactly two cells: $N_i \cap N_{i+1} = \{s_{i},\; s_{i+1}\}$, which are the updating cells themselves (see Figure~\ref{fig:overlap}).
We denote this overlap set $\mathcal{C}_{i}$. An ordered pair of PIs that resolves without contradiction at both overlap cells is a \emph{tile} (Definition~\ref{def:tile}); pairs whose enputs demand different values at any cell $s \in \mathcal{C}_{i}$ are excluded.
The remaining cells in the adjacent LUT entries, $s_{i-1}$ and $s_{i+2}$, each appear in only one of the two neighbourhoods and can be enputs only for their nearest updating cell.
The union $N_i \cup N_{i+1} = \{s_{i-1}, s_i, s_{i+1}, s_{i+2}\}$ spans both neighbourhoods. We call the concrete state of this union a \emph{tile window}, $\xi = (s_{i-1}, s_i, s_{i+1}, s_{i+2}) \in \{0,1\}^4$; it determines both $\eta_i$ and $\eta_{i+1}$. We refer to tile windows as \emph{windows} hereafter.

For each cell in $\mathcal{C}_i$, each PI places a demand, which can be either an enput ($0$ or $1$) or a wildcard ($\#$). 
These two demands must be resolved into a single consistent value. The overlap-resolution operator formalizes this:

%
%

\begin{definition}[Overlap resolution]
\label{def:res}
For two adjacent PIs $f'_a$ (left) and $f'_b$ (right), the operator $\sqcap\colon \{0,1,\#\}^{2} \to \{0,1,\#,\bot\}$ resolves their demands $(a, b)$ on a shared cell $s \in \mathcal{C}_i$, where $a$ is the demand placed by $f'_a$ and $b$ the demand placed by $f'_b$:
\begin{equation}
a \sqcap b \;=\;
\begin{cases}
\#    & a = \# \;\wedge\; b = \#, \\[2pt]
b     & a = \# \;\wedge\; b \in \{0,1\}, \\[2pt]
a     & a \in \{0,1\} \;\wedge\; b = \#, \\[2pt]
a     & a = b \in \{0,1\}, \\[2pt]
\bot  & a \neq b \;\wedge\; a,b \in \{0,1\}.
\end{cases}
\end{equation}
When $a \sqcap b \neq \bot$, the two demands are compatible; the resolved value is the unique consistent assignment. The operator returns the most specific value compatible with both demands, or $\bot$ when no compatible value exists.
\end{definition}

%
%

\begin{definition}[Tile]
\label{def:tile}
Let $f'_a = (a_\text{left}, a_\text{centre}, a_\text{right})$ and $f'_b = (b_\text{left}, b_\text{centre}, b_\text{right})$ be PIs resolving the transitions at adjacent cells $s_i$ and $s_{i+1}$, respectively. Their overlap set is $\mathcal{C}_i = \{s_i,\, s_{i+1}\}$, where $f'_a$ contributes entries $(a_\text{centre}, a_\text{right})$ and $f'_b$ contributes entries $(b_\text{left}, b_\text{centre})$. The ordered pair $(f'_a, f'_b)$ is a \emph{tile} if and only if the demands are compatible (Definition~\ref{def:res}) at both overlap cells: $a_\text{centre} \sqcap b_\text{left} \neq \bot$ and $a_\text{right} \sqcap b_\text{centre} \neq \bot$.
\end{definition}

%
%

\begin{definition}[Coupling weight]
\label{def:coupling-weight}
The \emph{coupling weight} $w$ of a tile $(f'_a, f'_b)$ counts the overlap cells at which both PIs place enput demands. At the first overlap cell $s_i$, the demands are $a_\text{centre}$ (from $f'_a$) and $b_\text{left}$ (from $f'_b$); at the second overlap cell $s_{i+1}$, they are $a_\text{right}$ and $b_\text{centre}$. An overlap cell contributes to $w$ when both demands are enputs ($\in \{0,1\}$). A tile with $w > 0$ is \emph{coupled}; a tile with $w = 0$ is \emph{decoupled}.
\end{definition}

\noindent(For the formal criterion and proof that direct coupling arises only at adjacent cells, see Supplementary Information, \S S1).

Only adjacent PIs can form tiles. Non-adjacent neighbourhoods ($N_i$ and $N_{i+2}$) share a single cell, $s_{i+1}$, which is not an updating cell for either PI; no coupling can arise. Any constraint at $s_{i+1}$ is resolved through sequential adjacent tiles, not direct coupling (see Supplementary Information, \S S1).

There are no three-cell or larger overlaps with coupled PIs at $r = 1$; higher-order constraints are resolved by sequential pairwise composition. This pairwise structure ensures that the composition is finite-state.

In valid tiles, each overlap cell $s \in \mathcal{C}_{i}$ receives a demand $a$ from $f'_a$ and a demand $b$ from $f'_b$, where each demand is an enput (0 or 1) or a wildcard (\#).
Three outcomes are possible at each overlap cell:

\begin{itemize}
\item \textbf{Coupled} ($a, b \in \{0,1\}$ and $a = b$): both PIs demand enputs and the values agree. The cell is a causal input to both transitions, coupling the two PIs at that cell.
\item \textbf{Absorbed} (one of $a, b \in \{0,1\}$, the other $\#$): only one PI demands an enput. The wildcard is absorbed; the enput constrains the cell but does not couple the PIs.
\item \textbf{Free} ($a = b = \#$): neither PI constrains the cell.
\end{itemize}

Tiles are ordered: the left and right PIs are distinguished, so $(f'_a, f'_b)$ and $(f'_b, f'_a)$ are distinct tiles. 
For a rule $\phi$ with $m$ prime implicants, at most $m^2$ ordered pairs exist; the contradiction-free subset (those that satisfy Definition~\ref{def:tile}) forms the \emph{tile catalogue}.
The tile catalogue is the complete static vocabulary of the rule's coupling repertoire: it lists every tile and its weight, from full coupling ($w = 2$) through partial coupling ($w = 1$) to decoupling ($w = 0$).
The catalogue is computable from the look-up table by exhaustive pairwise overlap resolution.

Each tile straddles the interface between two adjacent lattice sites. We call this interface a \emph{bond}, following the usage in lattice statistical mechanics~\cite{stauffer1994}. Each adjacent-cell pair in the lattice has one bond; each tile assigns a coupling weight to that bond. The tile catalogue therefore enumerates the coupling repertoire at every bond, paralleling the active/inactive distinction in bond percolation.

\emph{Scope of the coupling concept.} Throughout this paper, `coupling' and `decoupling' refer to pairwise coupling at the level of PI demands: whether two adjacent PIs share a cell in $\mathcal{C}_{i}$ that is simultaneously an enput in both. The coupling weight is a structural count derived from PI overlap. A $w = 0$ tile establishes the absence of such coupling; it does not preclude all statistical dependence or downstream influence between the two cells. The coupling weight does not quantify the magnitude of causal effect, information flow, or autonomy between cells.

\emph{Scope of the tile definition.} Definitions~\ref{def:tile} and~\ref{def:coupling-weight} are stated for ECAs: a one-dimensional lattice, radius $r = 1$, binary states.
Three features of this setting simplify the formalism.
First, the PI overlap set $\mathcal{C}_{i}$ contains exactly the two updating cells---so every shared enput is a direct coupling site (Supplementary Information, \S S1, Proposition~S1.3). 
Second, three consecutive radius-1 neighbourhoods do share the middle cell ($N_i \cap N_{i+1} \cap N_{i+2} = \{i{+}1\}$), but this does not create a primitive three-way direct coupling site; the constraint is fully mediated by sequential pairwise composition. The two-PI tile is therefore the only compositional unit required. Third, composition is sequential along a single spatial dimension; this is what makes the transducer (\S\ref{sec:tiling-transducer}) finite-state.

The conceptual core is lattice-independent: tile as contradiction-free PI pair, coupling weight as shared-enput count, and the three overlap outcomes (Coupled, Absorbed, Free; Supplementary Information, \S S1, Proposition~S1.2) apply wherever two Boolean functions share inputs. What changes at higher radii or in higher dimensions is the overlap geometry. When $|\mathcal{C}_{i}|$ grows, not all overlap cells need be updating cells and pairwise composition may no longer suffice ($r \geq 2$ permits three-way neighbourhood overlaps). Composition becomes a two-dimensional tiling problem rather than a one-dimensional scan. This paper develops the full machinery for the ECA case; the extension to general CAs is deferred to future work.

\textbf{Example.} Consider the tile $(f'_2, f'_2)$ for $\phi_{18}$, where $f'_2 = (\#, 1, \#)$. The left PI's centre and right cells are $(1, \#)$; the right PI's left and centre cells are $(\#, 1)$. At the first cell in $\mathcal{C}_{i}$: 1 (enput) meets \# (wildcard)---absorbed, resolved to~1. At the second: \# meets 1---absorbed, resolved to~1. Both overlaps resolve through absorption; no cell in $\mathcal{C}_{i}$ is an enput in both PIs. Weight $w = 0$: a decoupled tile. When this tile repeats across the lattice, it produces a chain of decoupled transitions---the domain. Now consider $(f'_3, f'_4)$ for $\phi_{18}$, where $f'_3 = (0, 0, 1)$ and $f'_4 = (1, 0, 0)$. The overlaps are $(0, 1)$ meeting $(1, 0)$: first cell 0 versus 1---contradicted. This pair is not a tile and never appears in the catalogue.

\subsection{The Tiling Transducer}
\label{sec:tiling-transducer}

The tile catalogue enumerates every pairwise coupling interaction but does not determine which tiles can compose sequentially: the overlap demands left unresolved by one tile constrain which tiles may follow, and these constraints propagate. The Tiling Transducer, $\mathcal{T}$, is a finite-state machine that resolves this sequential composition. It reads one PI per step, carries the unresolved overlap demands as state, and emits the resolved overlap value at each transition.

\begin{definition}[Tiling Transducer]
\label{def:tt}
$\mathcal{T} = (Q, \Sigma, \Gamma, \delta, \lambda, \mathbf{q}_0)$ is a finite-state transducer. The state set $Q \subseteq \{0, 1, \#\}^2$ consists of pairs $\mathbf{q} = (q_1, q_2)$: $q_1$ holds the resolved value from the most recent overlap resolution; $q_2$ holds the right entry of the most recently read PI, not yet resolved. The input alphabet $\Sigma = F'$ is the rule's PI set; the output alphabet $\Gamma = \{0, 1, \#\}$; the initial state $\mathbf{q}_0 = (\#, \#)$ imposes no inherited demand. At state $\mathbf{q} = (q_1, q_2)$, reading PI $f'_a = (a_\text{left}, a_\text{centre}, a_\text{right})$, compute the two overlap resolutions (Definition~\ref{def:res}): $o_1 = q_1 \sqcap a_\text{left}$ and $o_2 = q_2 \sqcap a_\text{centre}$. If either equals $\bot$, the transition is undefined. Otherwise:
\begin{align}
\delta\bigl(\mathbf{q},\; f'_a\bigr) &= (o_2,\; a_\text{right}), \\
\lambda\bigl(\mathbf{q},\; f'_a\bigr) &= o_1.
\end{align}
\end{definition}

The emission $o_1$ records whether the resolved overlap is an enput or causally free. The new state $(o_2, a_\text{right})$ carries the resolved overlap as the new $q_1$ and the raw right entry as the new $q_2$. The output alphabet is ternary: emissions of $0$ or $1$ mark enputs; an emission of $\#$ marks a position whose state is unconstrained by either adjacent PI. $\mathcal{T}$ is computable from the PI set and the overlap resolution operator alone; its construction requires no lattice configuration.

$\mathcal{T}$ is a grammar specifying which coupling sequences are structurally valid. Cycles in its state graph identify coupling patterns that can repeat indefinitely.

\emph{Lattice interpretation.} When $\mathcal{T}$ is applied to a concrete lattice row, $q_1$ corresponds to the resolved demand at the cell shared between the previous and current PIs; $q_2$ corresponds to the pending demand at the cell shared between the current and next PIs. Each $\#$-cell takes a concrete state at each time step, but that state plays no role in the causal structure of its neighbours. (For the full formal treatment, including the coupling weight per transition, the pruning criterion and the emission-to-domain bridge, see Supplementary Information, \S S2).

\begin{definition}[Effective tiling transducer]
\label{def:effective-tt}
Multiple PIs may be compatible with a given state $\mathbf{q}$, each placing different enput demands. A transducer edge is \emph{effective} if its PI imposes the fewest demands (has the highest wildcard count) among those compatible with at least one concrete neighbourhood $\eta_i$ consistent with $\mathbf{q}$ (Definition~S2.1 in Supplementary Information, \S S2). The effective tiling transducer $\mathcal{T}_{\mathrm{eff}}$ is the subgraph of $\mathcal{T}$ restricted to effective edges; it represents the maximally canalizing reading of every state.
\end{definition}

\emph{Scope of the compression.} $\mathcal{T}_{\mathrm{eff}}$ is a specific compression: the maximally canalizing reading of each state. We do not claim that this compression is optimal with respect to a general predictive-cost or state-space-minimization objective. It is the natural representation for the structural coupling diagnostics developed here (direct coupling, forced decoupling and wildcard-emission cycles) because these diagnostics are defined at the level of prime implicant demands. Alternative coarse-grainings, optimized for different observables, may compress the transducer differently.

\emph{Nondeterminism at ties.} When two or more PIs are effective (Definition~\ref{def:effective-tt}) at a given state $\mathbf{q}$, all appear as parallel outgoing edges. $\mathcal{T}_{\mathrm{eff}}$ is deterministic where one PI dominates and nondeterministic at ties. At a tie, the coupling at that bond spans the range across all parallel edges, a form of positional redundancy.

\begin{figure}[!ht]
\centering
\includegraphics[width=0.45\columnwidth]{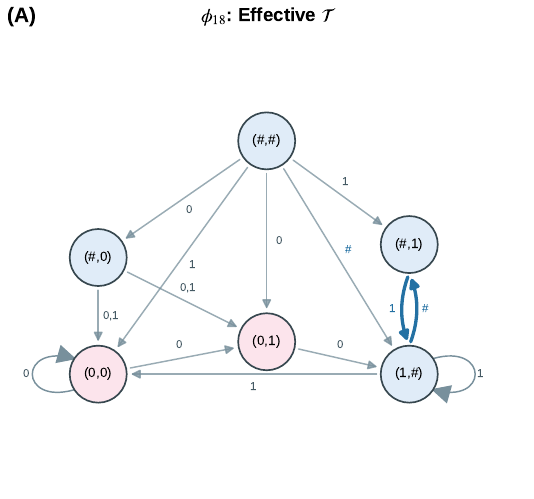}
\hfill
\includegraphics[width=0.45\columnwidth]{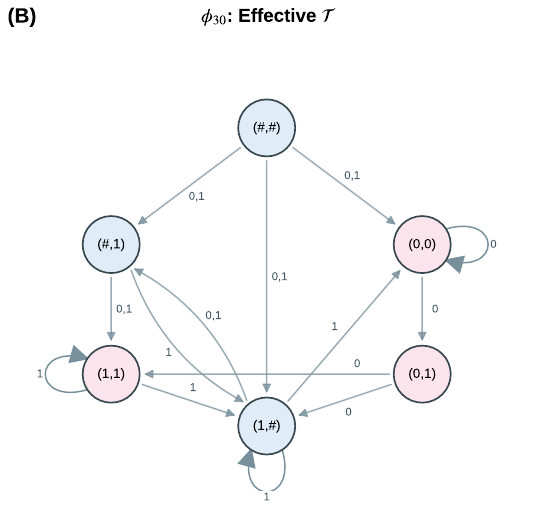}
\caption{\label{fig:tt} Effective tiling transducer state graphs. (A)~$\phi_{18}$ (6 states, 16 transitions). The wildcard-emission cycle between $(1, \#)$ and $(\#, 1)$ is highlighted in blue. (B)~$\phi_{30}$ (6 states, 19 transitions, $\mathcal{T}_{\text{eff}} = \mathcal{T}$). There are no wildcard-emission cycles. Blue nodes contain at least one wildcard component; pink nodes have both components binary. Edge labels show emitted values; multi-valued labels ($0{,}1$) indicate parallel edges---different effective PIs that connect the same state pair, each emitting a distinct $\lambda$. Parallel edges that cover the same concrete neighbourhood are ties (nondeterminism); those that cover different neighbourhoods are resolved by the input.
\emph{Reading the graph: a two-step trace through $\phi_{18}$.} PIs are listed in Table~\ref{tab:PIs}. Start at $\mathbf{q} = (\#, \#)$ and read $f'_0 = (0, \#, 0)$. The two overlap resolutions are $o_1 = q_1 \sqcap a_\text{left} = \# \sqcap 0 = 0$ and $o_2 = q_2 \sqcap a_\text{centre} = \# \sqcap \# = \#$. The transducer emits $\lambda = o_1 = 0$ and moves to $\delta = (o_2,\, a_\text{right}) = (\#, 0)$: no constraint inherited at the first overlap, a pending demand of~0 at the second. Both overlaps involve a wildcard, so $w = 0$ (a decoupled bond).
Now read $f'_3 = (0, 0, 1)$ from $(\#, 0)$. The first overlap: $\# \sqcap 0 = 0$; emission $\lambda = 0$. The second: $0 \sqcap 0 = 0$. Both the pending demand from $f'_0$ and the centre entry of $f'_3$ are enputs, and they agree. This overlap is coupled ($w = 1$): the value~0 at that cell is a causal input to both transitions. The new state is $\delta = (0, 1)$. Two steps, two bonds: the first decoupled, the second coupled. The graph encodes this distinction for every PI sequence the rule permits.}
\end{figure}


\subsection{The three-way bond classification}
\label{sec:three-way}

The transducer composes tiles sequentially, carrying unresolved demands as state. Its cycles identify coupling patterns that can repeat indefinitely in the input sequence. But at a given state $\mathbf{q}$, multiple PIs may be effective---parallel edges in $\mathcal{T}_{\mathrm{eff}}$---and these may produce tiles with different coupling weights. The transducer records that both coupled and decoupled transitions are structurally valid from that state; it does not settle which the bond must be. A finer question arises: what is the coupling status at a bond---is coupling between adjacent transitions guaranteed, possible, or structurally excluded?

Two or more PIs may cover a given neighbourhood $\eta_i$. Among these, those with the highest wildcard count are \emph{maximally canalizing}: they impose the fewest enput demands while still determining the output. The set of maximally canalizing PIs for $\eta_i$ forms the \emph{option set} for that neighbourhood---the same criterion that defines effective edges in $\mathcal{T}_{\mathrm{eff}}$ (Definition~\ref{def:effective-tt}), now applied to a fully specified neighbourhood rather than a transducer state. A tile window $\xi \in \{0,1\}^4$ (\S\ref{sec:tiles}) fixes both adjacent neighbourhoods and hence determines two option sets. The Cartesian product of these two sets generates all applicable tiles at that bond, and the range of their coupling weights settles the coupling status.

\begin{definition}[Three-way bond classification]
\label{def:three-way}
Let $L$ and $R$ be the option sets for $\eta_i$ and $\eta_{i+1}$ respectively, and let $W = \{w(f'_a, f'_b) : (f'_a, f'_b) \in L \times R\}$ be the set of coupling weights (Definition~\ref{def:coupling-weight}) across all tiles. The bond is:
\begin{enumerate}
\item \textbf{Forced-coupled} if $\min W > 0$: every tile shares at least one enput in the overlap. No maximally canalizing PI assignment eliminates the coupling.
\item \textbf{Optional} if $\min W = 0$ and $\max W > 0$: some tiles share enputs; others do not. The coupling is present under one maximally canalizing PI assignment and absent under another.
\item \textbf{Forced-decoupled} if $\max W = 0$: no tile shares an enput in the overlap. The adjacent transitions are directly decoupled under every maximally canalizing PI assignment.
\end{enumerate}
\end{definition}

\textbf{The bond-classification map.}\label{par:bond_map}
The classification is deterministic: the window $\xi$ fixes both cells' neighbourhoods, both outputs (via $\phi$), both PI option sets, and hence the bond class. The \emph{bond-classification map}
\begin{equation}\label{eq:BR}
  B_\phi : \{0,1\}^4 \;\longrightarrow\; \{C, O, D\}
\end{equation}
assigns each window its bond class. $B_\phi(\xi)$ is computable from the look-up table alone: no transducer run or lattice configuration is needed. Since there are only $2^4 = 16$ possible windows, $B_\phi$ can be evaluated exhaustively. The resulting triple of counts---how many windows map to $C$, $O$ and $D$---is the rule's \emph{window signature}.

\begin{table}[tbp]
\centering
\begin{tabular}{lccc}
\toprule
\textit{Rule} & \textit{Forced-coupled} & \textit{Optional} & \textit{Forced-decoupled} \\
\midrule
$\phi_{18}$ & 11 & 0 & 5 \\
$\phi_{30}$ & 12 & 4 & 0 \\
\bottomrule
\end{tabular}
\caption{\label{tab:windows} Window signatures for $\phi_{18}$ and $\phi_{30}$}
\end{table}

Note that $\phi_{18}$ has 5 forced-decoupled windows and zero optional. At those configurations, no maximally canalizing PI assignment couples the adjacent transitions. Compare with $\phi_{30}$, which has zero forced-decoupled windows. No configuration guarantees decoupling. Its four optional windows mean that some bonds are decoupled under one maximally canalizing PI assignment but coupled under another. Whether this asymmetry carries predictive significance is an empirical question; the next two subsections develop the spatial and temporal conditions under which it could.


\subsection{Wildcard-emission cycles and self-sustaining decoupling}
\label{sec:wildcard-cycles}

A cycle in $\mathcal{T}_{\mathrm{eff}}$ identifies a coupling pattern that can repeat indefinitely in the PI sequence. Not all cycles are equivalent. Some emit only enputs ($0$ or $1$); others emit at least one wildcard ($\#$). A cell for which the transducer emits $\#$ is one whose state is not an enput in either adjacent effective PI (Definition~\ref{def:tt}). The cell is thus \emph{causally free}: it plays no role in the causal determination of the adjacent transitions. The distinction between enput-only and wildcard-emitting cycles determines whether a rule's coupling repertoire includes repeating decoupled structure.

\begin{definition}[Wildcard-emission cycle]
\label{def:wildcard-cycle}
A cycle in $\mathcal{T}_{\mathrm{eff}}$ is a \emph{wildcard-emission cycle} if at least one edge in the cycle satisfies $\lambda(\mathbf{q}, f'_a) = \#$.
\end{definition}

A wildcard-emission cycle of length $k$ with $m$ wildcard emissions per traversal produces a repeating PI sequence: each traversal generates $k$ consecutive tiles containing $m$ causally free cells, and the same sequence of coupling weights and emissions recurs with every traversal. The entire construction is readable from the look-up table alone. Whether the repeating pattern persists under the update rule is a separate question (\S\ref{sec:self-consistency}).

\textbf{From emissions to domain language.} The emission alphabet is ternary: $0$, $1$ and $\#$. Emissions encode \emph{causal status}, not cell values: an enput ($0$ or $1$) marks a position whose output the PI determines; a wildcard ($\#$) marks a position free to take either value. To obtain the cell-value pattern that a cycle generates, each emission is mapped through the output of the PI that produced it. An enput at a position where the PI outputs $v$ maps to $v$; a wildcard maps to $\Sigma$ (either binary value). This PI-output mapping converts the causal-status sequence into a cell-value regular expression---the candidate domain language. The mapping is computable from the look-up table; the framework derives the correspondence without examining the dynamics.

\textbf{The wildcard-emission cycle in $\phi_{18}$.} $\mathcal{T}_{\mathrm{eff}}$ has a two-state cycle. State $(1, \#)$ transitions to $(\#, 1)$ via $f'_1 = (1, \#, 1)$, emitting 1; then $(\#, 1)$ transitions back to $(1, \#)$ via $f'_2 = (\#, 1, \#)$, emitting \#. The emission sequence is $1, \#, 1, \#, \ldots$, alternating enputs and causally free cells. Applying the PI-output mapping: both cycle PIs output~0, so enputs map to~0 and wildcards to~$\Sigma$---yielding $(0\Sigma)^*$, the regular domain that computational mechanics identifies for $\phi_{18}$~\cite{hanson1997}.
What does the cycle reveal? $f'_2 = (\#, 1, \#)$ has wildcards at left and right; when composed through $\mathcal{T}_{\mathrm{eff}}$, those wildcards become emissions---cells that the next PI inherits as unconstrained inputs. These are exactly the positions where $f'_1 = (1, \#, 1)$ demands no constraint: its wildcard sits at the centre. Each PI in the cycle feeds the preconditions of the next---$f'_2$ produces causally free cells that $f'_1$ absorbs, and $f'_1$ produces the enput that $f'_2$ requires. The cycle is structurally self-feeding, but whether this property translates into dynamical persistence is the question that self-consistency (\S\ref{sec:self-consistency}) addresses.

\textbf{There are no wildcard-emission cycles in $\phi_{30}$.} Every cycle in the $\mathcal{T}_{\mathrm{eff}}$ of $\phi_{30}$ emits only enputs. The wildcards in $\phi_{30}$ sit at centre or right positions; when composed through $\mathcal{T}_{\mathrm{eff}}$, they enter the state but never surface as emissions. Wildcard positions appear locally but do not chain into a repeating decoupled structure. The window signature (\S\ref{sec:three-way}) confirms this from the other direction: zero forced-decoupled windows. The two diagnostics---no wildcard-emission cycle in $\mathcal{T}_{\mathrm{eff}}$, no forced-decoupled class in $B_\phi$---converge: $\phi_{30}$ lacks every structural condition that $\phi_{18}$ exhibits. What this absence implies for the dynamics of $\phi_{30}$ is examined in the case studies (Section~\ref{sec:case-studies}).


\subsection{Self-consistency: temporal closure}
\label{sec:self-consistency}

The wildcard-emission cycle (\S\ref{sec:wildcard-cycles}) identifies a repeating PI composition whose emissions include causally free cells. The bond-classification map $B_\phi$ (\S\ref{sec:three-way}) classifies each window's bond from the coupling weights of its most-general PI pair; for some rules, this classification includes forced-decoupled windows. These are independent structural diagnostics: the first captures how PIs compose across space, the second captures how adjacent PIs constrain one another at each bond. Neither addresses time. A compositional pattern at time~$t$ may produce an entirely different bond structure at $t+1$---the pattern would then be structurally permitted but dynamically transient. Self-consistency closes this gap. It asks whether a window classified~$D$ at time~$t$ remains~$D$ at $t+1$.

\textbf{The past-lightcone constraint.} The successor of a window $\xi^{t} = (s_{i-1}, s_i, s_{i+1}, s_{i+2})$ at time $t+1$ is $\xi^{t+1} = (s^{t+1}_{i-1}, s^{t+1}_i, s^{t+1}_{i+1}, s^{t+1}_{i+2})$, where each $s^{t+1}_j = \phi(s_{j-1}, s_j, s_{j+1})$. Computing these four outputs requires six cells $s_{i-2}, \ldots, s_{i+3}$ at time $t$, spanning three consecutive windows: $\xi_L = (s_{i-2}, s_{i-1}, s_i, s_{i+1})$, $\xi_C = (s_{i-1}, s_i, s_{i+1}, s_{i+2})$ and $\xi_R = (s_i, s_{i+1}, s_{i+2}, s_{i+3})$.

If $D$-classified bonds are to persist in a contiguous region, they must do so when flanked by other $D$-classified bonds. A \emph{forced-decoupled triple} is a configuration in which all three consecutive windows are classified $D$ by $B_\phi$. Self-consistency asks whether the central bond remains $D$ after one application of $\phi$.

\begin{definition}[Self-consistency]
\label{def:self-consistency}
Let $(\xi_L, \xi_C, \xi_R)$ be a forced-decoupled triple: three consecutive windows with $B_\phi(\xi_L) = B_\phi(\xi_C) = B_\phi(\xi_R) = D$. Denote the six underlying cells $s_{i-2}, \ldots, s_{i+3}$. The \emph{successor central window} is
\begin{equation}\label{eq:succ-window}
  \xi^{t+1} = \bigl(\phi(s_{i-2}, s_{i-1}, s_i),\; \phi(s_{i-1}, s_i, s_{i+1}),\; \phi(s_i, s_{i+1}, s_{i+2}),\; \phi(s_{i+1}, s_{i+2}, s_{i+3})\bigr).
\end{equation}
The triple is \emph{self-consistent} if $B_\phi(\xi^{t+1}) = D$. A rule is self-consistent if every forced-decoupled triple is self-consistent.
\end{definition}

\emph{Scope.} Self-consistency is a one-step, classification-level check. It asks whether the bond class $D$ is preserved, not whether the same PIs, tiles, or cell values recur. It catches period-1 stability only; periodic forced-decoupled patterns with period $k > 1$ would require $k$-fold composition and are beyond the present scope. The check is exhaustive over $\{0,1\}^6$: every possible 6-bit input is tested. No lattice simulation is required.

For $\phi_{18}$, all forced-decoupled triples are self-consistent. For $\phi_{30}$, the question is vacuous: there are no forced-decoupled windows and hence no triples to test.


\subsection{Structural hypothesis}
\label{sec:hypothesis}

The framework developed in Sections~\ref{sec:tiles}--\ref{sec:self-consistency} derives three constructs from the look-up table alone: the bond-classification map $B_\phi$ and its window signature (\S\ref{sec:three-way}), the wildcard-emission cycle (\S\ref{sec:wildcard-cycles}), and the self-consistency check (\S\ref{sec:self-consistency}). For $\phi_{18}$, these converge: five forced-decoupled windows, a wildcard-emission cycle whose emission sequence corresponds, through PI outputs, to the known regular domain, and self-consistency across all forced-decoupled triples. For $\phi_{30}$, they converge in the opposite direction: zero forced-decoupled windows, no wildcard-emission cycle, and a rule that resists domain--particle decomposition.

Two cases do not establish a general law, but they suggest a structural hypothesis:

\emph{A rule forms emergent self-sustaining domains only if (i) $\mathcal{T}_{\mathrm{eff}}$ contains a wildcard-emission cycle, (ii) $B_\phi$ classifies at least one window as forced-decoupled ($n_\text{fd} \geq 1$), and (iii) all forced-decoupled triples are self-consistent.} Constitutive decoupling (Level~1 in the hierarchy of \S\ref{sec:hierarchy}) produces self-sustaining decoupling without a cycle; the conjunction targets the non-trivial case. The set of forced-decoupled windows, $S_\phi = \{\xi : B_\phi(\xi) = D\}$, and its size $n_\text{fd} = |S_\phi|$, quantify the structural support available for domain formation. The 88-rule classification (\S\ref{sec:hierarchy}) uses only conditions~(i) and~(ii) as coarse diagnostics; condition~(iii) further subdivides the 46-rule emergent class (Supplementary Information, \S S9).

\noindent The hypothesis is testable. Sections~\ref{sec:case-studies} and~\ref{sec:hierarchy} examine $\phi_{18}$ and $\phi_{30}$ in detail and extend the classification to all 88 ECA equivalence classes. Section~\ref{sec:dynamical-validation} tests the quantitative prediction that $n_\text{fd}$ correlates with the dynamical prevalence of forced-decoupled bonds.



\section{Case studies: $\phi_{18}$ and $\phi_{30}$}
\label{sec:case-studies}

\begin{table}[!ht]
\centering
\begin{tabular}{llccc}
\toprule
& \textit{Name} & \textit{Schema} & \textit{Output} & $|\cdot|_\#$ \\
\midrule
\multirow{5}{*}{\textbf{$\phi_{18}$}} & $f'_0$ & $(0, \#, 0)$ & 0 & 1 \\
& $f'_1$ & $(1, \#, 1)$ & 0 & 1 \\
& $f'_2$ & $(\#, 1, \#)$ & 0 & 2 \\
& $f'_3$ & $(0, 0, 1)$ & 1 & 0 \\
& $f'_4$ & $(1, 0, 0)$ & 1 & 0 \\
\midrule
\multirow{6}{*}{\textbf{$\phi_{30}$}} & $f'_0$ & $(0, 0, 0)$ & 0 & 0 \\
& $f'_1$ & $(1, \#, 1)$ & 0 & 1 \\
& $f'_2$ & $(1, 1, \#)$ & 0 & 1 \\
& $f'_3$ & $(0, \#, 1)$ & 1 & 1 \\
& $f'_4$ & $(0, 1, \#)$ & 1 & 1 \\
& $f'_5$ & $(1, 0, 0)$ & 1 & 0 \\
\bottomrule
\end{tabular}
\caption{\label{tab:PIs} Prime implicants of $\phi_{18}$ and $\phi_{30}$}
\end{table}

The structural hypothesis (\S\ref{sec:hypothesis}) rests on two cases. $\phi_{18}$ and $\phi_{30}$ produce aperiodic, apparently random spacetime, yet the tiling framework identifies fundamentally different causal architectures from their look-up tables. This section applies the full diagnostic chain: $\mathcal{T}_{\mathrm{eff}}$, bond-classification map, wildcard-emission cycle, and self-consistency to each of these ECAs, derives structural predictions, and tests them against spacetime dynamics data. Exploratory simulations used $N = 101$ cells, $T = 200$ steps, random initial configuration (seed 42). Every bond was classified via the three-way classification using option-aware annotation that tracks the full PI option set at each cell.

\subsection{$\phi_{18}$: rigid topology, self-sustaining decoupling}

$\phi_{18}$ has five prime implicants (Table~\ref{tab:PIs}). $\mathcal{T}_{\mathrm{eff}}$ has 6 states and 16 transitions, reduced from 8 states and 23 transitions in the full $\mathcal{T}$; the pruning removes transitions where a less-general PI is subsumed by a more general one at the same state. The structural predictions from $\mathcal{T}_{\mathrm{eff}}$ are:

\begin{enumerate}

\item{Wildcard-emission cycle.} $\mathcal{T}_{\mathrm{eff}}$ contains a two-state cycle $[(1, \#) \to (\#, 1) \to (1, \#)]$ with emission pattern $1, \#, 1, \#, \ldots$---alternating enputs and causally free cells. Through the PI-output mapping (\S\ref{sec:wildcard-cycles}), this emission sequence corresponds to the $(0\Sigma)^*$ domain.

\item{Self-sustaining decoupled bonds.} $B_\phi$ classifies 5 of 16 windows as forced-decoupled. All 5 forced-decoupled triples are self-consistent (\S\ref{sec:self-consistency}). The $D$ classification should persist across time steps.

\item{Rigid segmentation.} The window signature (11, 0, 5) has zero optional windows: $B_\phi(\xi) \in \{C, D\}$ for every $\xi$. No bond should be ambiguous.

\item{Narrow vocabulary.} The pruning from 23 to 16 transitions means only 16 of 25 possible PI-pair types are structurally valid in context. The spacetime should use a restricted pairing vocabulary.

\end{enumerate}

\begin{figure}[tbp]
\centering
\includegraphics[width=0.7\columnwidth]{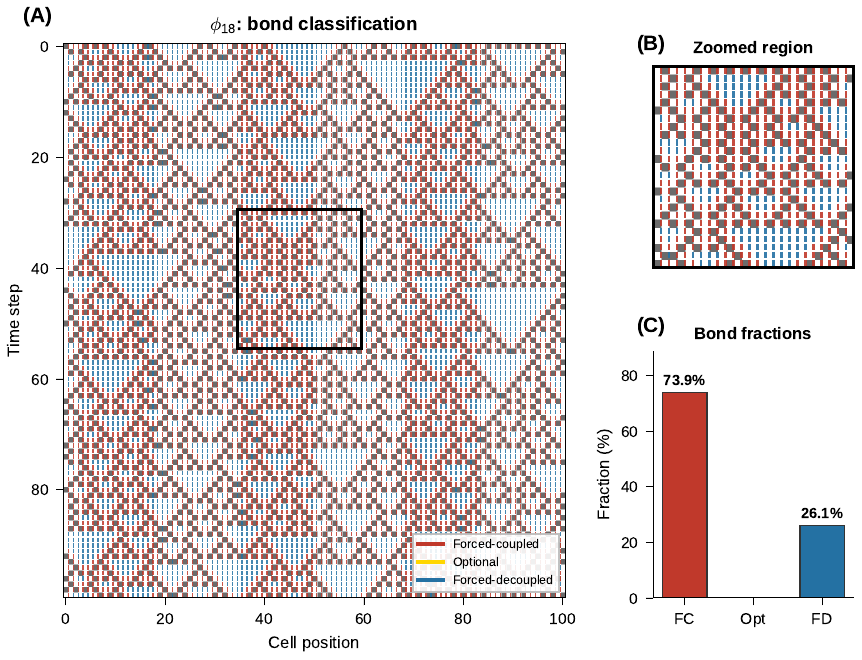}
\caption{\label{fig:r18} $\phi_{18}$ bond classification ($N = 101$, $T = 200$, seed 42, transient $= 50$). (A)~Full spacetime with bond overlay---forced-coupled in red, forced-decoupled in blue. The $(0\Sigma)^*$ domain appears as contiguous blue regions; particles appear as red threads. Black rectangle marks the zoomed region. (B)~$25 \times 25$ detail showing the alternating coupled/decoupled micro-structure. (C)~Global bond fractions: 73.9\% forced-coupled, 26.1\% forced-decoupled, 0\% optional.}
\end{figure}

\textbf{Results.} Post-transient bonds are 73.9\% forced-coupled, 26.1\% forced-decoupled and 0\% optional (see Figure~\ref{fig:r18}). The forced-decoupled fraction matches the domain: alternating cells of the $(0\Sigma)^*$ pattern produce forced-decoupled bonds wherever two output-0 PIs with wildcards sit adjacent. The domain pair $f'_0|f'_0$ accounts for 25.4\% of all post-transient pair observations and appears at all 150 post-transient steps. Only 9 of 25 possible PI-pair types appear in the spacetime.

\textbf{What the framework makes legible.} Three diagnostics converge. The wildcard-emission cycle captures the compositional pathway: $f'_2$ produces causally free cells that $f'_1$ absorbs, and the repeating composition corresponds, through the PI-output mapping (\S\ref{sec:wildcard-cycles}), to the $(0\Sigma)^*$ domain. $B_\phi$ classifies five windows as forced-decoupled; self-consistency confirms that the $D$ classification persists under~$\phi$. The remaining PIs---$f'_3 = (0, 0, 1)$ and $f'_4 = (1, 0, 0)$, both with zero wildcards---appear as fully coupled chains threading through the decoupled background.

\subsection{$\phi_{30}$: flexible topology, no forced decoupling}

$\phi_{30}$ has six prime implicants (Table~\ref{tab:PIs}). $\mathcal{T}_{\mathrm{eff}}$ equals the full $\mathcal{T}$: 6 states, 19 transitions, zero pruning. Two neighbourhoods---$(0, 1, 1)$ and $(1, 1, 0)$---have tied PIs of equal wildcard count, producing the four optional windows in the signature. The structural predictions from $\mathcal{T}_{\mathrm{eff}}$ are:

\begin{enumerate}
\item{No self-sustaining decoupled structure.} No wildcard-emission cycle exists. Causal freedom cannot chain across cells. No persistent decoupled structure should appear.

\item{Zero forced-decoupled bonds.} The window signature is (12, 4, 0). No four-cell configuration forces decoupling. Every bond that \emph{can} be described as decoupled also admits a coupled description of equal generality.

\item{Full vocabulary.} Zero pruning means all structurally valid PI-pair types should appear in the spacetime.

\end{enumerate}

\begin{figure}[tbp]
\centering
\includegraphics[width=0.7\columnwidth]{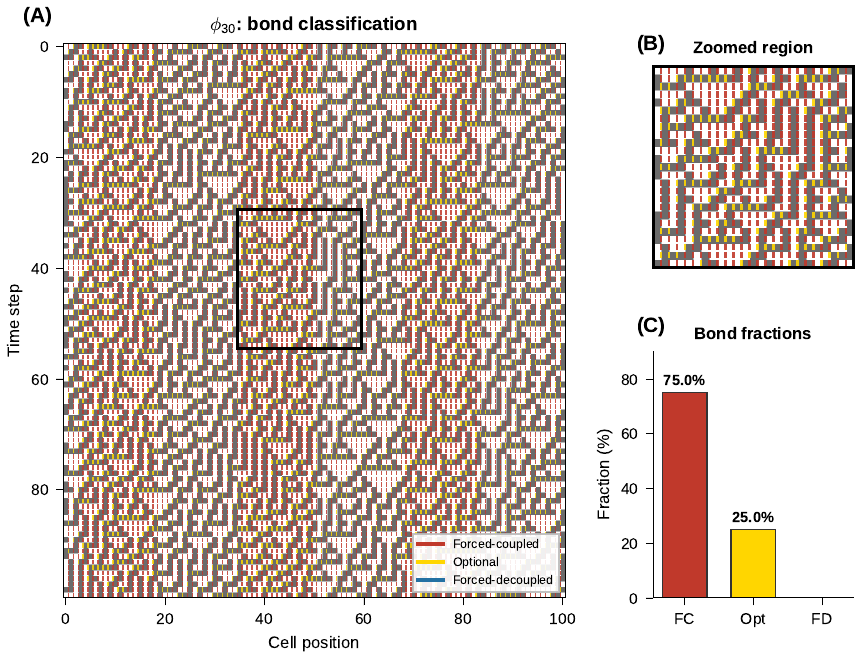}
\caption{\label{fig:r30} $\phi_{30}$ bond classification ($N = 101$, $T = 200$, seed 42, transient $= 50$). (A)~Full spacetime with bond overlay---forced-coupled in red, optional in yellow. No blue (forced-decoupled) regions appear. Black rectangle marks the zoomed region. (B)~$25 \times 25$ detail showing the uniformly coupled/optional micro-structure. (C)~Global bond fractions: 75.0\% forced-coupled, 25.0\% optional, 0\% forced-decoupled.}
\end{figure}

\textbf{Results.} Post-transient bonds are 75.0\% forced-coupled, 25.0\% optional and 0\% forced-decoupled (see Figure~\ref{fig:r30}). Zero forced-decoupled bonds appear across 15,150 post-transient observations. Optional bonds appear but never persist---every optional bond becomes forced-coupled at the next step. All 18 compatible PI-pair types appear in the spacetime with near-uniform frequency.

\textbf{What the framework makes legible.} The PIs of $\phi_{30}$ contain wildcards, but no window-level configuration composes them into forced decoupling: $B_{\phi_{30}}(\xi) \in \{C, O\}$ for every $\xi$, so $S_{\phi_{30}} = \varnothing$ and $n_\text{fd} = 0$. Causal freedom appears locally but $\mathcal{T}_{\mathrm{eff}}$ cannot chain it across cells: wildcards sit at centre or right positions and enter the transducer state rather than surfacing as emissions. None of the three conditions in the structural hypothesis (\S\ref{sec:hypothesis}) is met.

\subsection{Structural comparison}

Both rules produce aperiodic spacetime that resists visual classification. The framework identifies different causal architectures from their LUTs (Table~\ref{tab:contrast}). The three conditions of the structural hypothesis (\S\ref{sec:hypothesis})---wildcard-emission cycle, $n_\text{fd} \geq 1$, and self-consistency---are all satisfied for $\phi_{18}$ ($n_\text{fd} = 5$, all triples self-consistent) and none for $\phi_{30}$ ($n_\text{fd} = 0$, no cycle). The dynamical measurements confirm each prediction: $\phi_{18}$ produces persistent forced-decoupled bonds; $\phi_{30}$ produces exactly zero. These exploratory results establish that the framework's structural diagnostics, derived entirely from the look-up table, correctly separate two rules with opposite coupling architectures. Whether this separation generalizes requires a systematic test across the full ECA rule space. Section~\ref{sec:hierarchy} applies the diagnostic chain to all 88 equivalence classes; Section~\ref{sec:dynamical-validation} tests the predictions against spacetime simulations.

\begin{table}[!ht]
\centering
\begin{tabular}{lccc}
\toprule
\textit{Property} & \textit{$\phi_{18}$} & \textit{$\phi_{30}$} \\
\midrule
PIs & 5 & 6 \\
Canalization asymmetry & $4 : 0$ & $2 : 2$ \\
$\mathcal{T}_{\mathrm{eff}}$ states & 6 & 6 \\
Pruned transitions & 7 & 0 \\
Ties & 0 & 2 \\
Wildcard-emission cycle & Yes (2-state) & No \\
Window signature & (11, 0, 5) & (12, 4, 0) \\
Self-consistent triples & 5/5 & --- \\
Dynamical $\theta_\text{fd}$ & 26.1\% & 0.0\% \\
Mean FD run length & 3.5 & 0.0 \\
\bottomrule
\end{tabular}
\caption{\label{tab:contrast} Structural comparison of $\phi_{18}$ and $\phi_{30}$}
\end{table}


\section{Structural classification of all ECA rules}
\label{sec:hierarchy}

The structural hypothesis (\S\ref{sec:hypothesis}) was formulated from two cases. Testing it requires evaluating the diagnostic chain across all 256 ECA rules. Every ECA rule is equivalent under reflection and input--output complement to one of 88 representatives~\cite{wolfram2002}, so the 88 equivalence classes exhaust the rule space. The three structural diagnostics---wildcard-emission cycle, window signature, and self-consistency---are computable from the LUT alone. We evaluated all three for each representative.

\subsection{Results}

Two binary diagnostics concerning the presence of a wildcard-emission cycle, and whether $n_\text{fd} = 0$, produce a $2 \times 2$ partition. Table~\ref{tab:class} shows the result for the 88 representative ECAs.
The largest group (46 rules, 52.3\%) satisfies both conditions of the structural hypothesis: a wildcard-emission cycle and at least one forced-decoupled window. Of these, 41 have $n_\text{fd} \geq 2$. The 22 rules with $n_\text{fd} = 0$ lack any structural basis for forced-decoupled bonds. The remaining 20 rules present an unexpected pattern: forced-decoupled windows exist ($n_\text{fd} \geq 1$) but no wildcard-emission cycle forms---how can decoupling arise without the spatial mechanism? A fourth group---11 rules with a cycle but $n_\text{fd} = 0$---poses the converse: can a cycle produce forced decoupling without any forced-decoupled windows?

\begin{table}[tbp]
\centering
\begin{tabular}{lcc}
\toprule
 & $n_\text{fd} \geq 1$ & $n_\text{fd} = 0$ \\
\midrule
Wildcard-emission cycle & 46 & 11 \\
No wildcard-emission cycle & 20 & 11 \\
\bottomrule
\end{tabular}
\caption{\label{tab:class} Partition of the 88 ECA equivalence classes by two binary structural diagnostics: presence of a wildcard-emission cycle in $\mathcal{T}_{\mathrm{eff}}$, and existence of at least one forced-decoupled window ($n_\text{fd} \geq 1$). The 46 rules satisfying both coarse diagnostics are candidates for emergent decoupling; the finer self-consistency criterion further subdivides them (Supplementary Information, \S S9). The 22 rules with $n_\text{fd} = 0$ lack any structural basis for forced-decoupled bonds.}
\end{table}

\subsection{Two special cases}

\textbf{Forced-decoupled windows without a cycle.} Twenty rules have $n_\text{fd} \geq 1$ but no wildcard-emission cycle. Six of these ($\phi_{0}$, $\phi_{15}$, $\phi_{51}$, $\phi_{90}$, $\phi_{170}$, $\phi_{204}$) have all 16 windows forced-decoupled. These are constant functions, identity, complement and XOR---rules whose outputs depend on at most one or two inputs. Every PI individually contains wildcards, so every bond is decoupled by PI structure alone; no spatial composition is needed. The wildcard-containing states in $\mathcal{T}_{\mathrm{eff}}$ are transient: the transducer passes through them but never returns, so no cycle forms. The remaining 14 rules in this group have $n_\text{fd}$ between 1 and 6. We call this mode \emph{constitutive decoupling}: independence is built into each PI rather than emerging from how PIs compose.

\textbf{A cycle but no forced-decoupled windows.} Eleven rules have wildcard-emission cycles but $n_\text{fd} = 0$. The cycle's wildcard emissions occur at specific $\mathcal{T}_{\mathrm{eff}}$ edges, but the window-first analysis---which considers all effective PIs at each four-cell window---finds that at every window, at least one PI choice forces coupling. The decoupling is \emph{optional}, not forced: $\mathcal{T}_{\mathrm{eff}}$ has a path that produces decoupled behaviour, but an alternative PI of equal generality always exists that couples the bond.

\subsection{Three-level hierarchy}

The four categories reflect three structurally distinct modes of decoupling. Decoupling is either built into each PI individually (constitutive), emergent through $\mathcal{T}_{\mathrm{eff}}$ composition, or absent. The hierarchy is computable from the look-up table alone; whether it predicts dynamical behaviour is the subject of Section~\ref{sec:dynamical-validation}.

\textbf{Constitutive decoupling} (20 rules). Decoupling is a property of each PI, not of how PIs compose. No spatial mechanism is needed and no cycle forms.

\textbf{Emergent decoupling} (46 rules, including $\phi_{18}$). The wildcard-emission cycle provides the spatial mechanism; forced-decoupled windows provide the structural basis. Within this group, $n_\text{fd}$ ranges from 1 to 16, offering a quantitative structural gradient. The finer self-consistency breakdown is reported in the Supplementary Information (Table~S10, \S S9).

\textbf{No forced decoupling} (22 rules, including $\phi_{30}$ and $\phi_{110}$). Eleven have wildcard-emission cycles with only optional decoupling; eleven have neither a cycle nor forced-decoupled windows.

The hierarchy yields two diagnostic questions for any ECA rule: (1) Does $\mathcal{T}_{\mathrm{eff}}$ contain a wildcard-emission cycle? (2) Does $B_\phi$ classify at least one window as forced-decoupled? Where both answers are yes, the framework predicts persistent forced-decoupled bonds, with $n_\text{fd}$ as the quantitative structural predictor.


\section{Dynamical validation}
\label{sec:dynamical-validation}

The structural classification predicts which rules produce forced-decoupled bonds and how many. This section tests those predictions against spacetime simulations of all 88 equivalence classes.

\subsection{Method}

For each rule, we simulated 50 random initial configurations ($N = 101$ cells, $T = 200$ steps, seeds 42--91, periodic boundaries). At each post-transient time step ($t \geq 50$), we assigned the most-general PI(s) to each cell based on the actual neighbourhood values and classified each bond via the three-way classification (\S\ref{sec:three-way}). The primary metric is the \emph{forced-decoupled fraction} $\theta_\text{fd}$: the fraction of post-transient bonds classified as forced-decoupled, averaged over seeds.

\subsection{The support/occupancy identity}
\label{sec:identity}

The structural support $S_\phi$ and its size $n_\text{fd}$ (\S\ref{sec:hypothesis}) partition the 16 possible windows into those that guarantee forced decoupling and those that do not. On a concrete lattice row $\mathbf{x}$, each window $\xi$ occurs with empirical frequency $\mu_{\mathbf{x}}(\xi)$. The forced-decoupled fraction of that row is the total frequency mass on $S_\phi$:
\begin{equation}\label{eq:fd_identity}
  \theta_\text{fd}(\mathbf{x}) \;=\; \sum_{\xi \in \{0,1\}^4} \mu_{\mathbf{x}}(\xi)\;\mathbf{1}[B_\phi(\xi) = D].
\end{equation}

This identity decomposes $\theta_\text{fd}$ into two factors: the indicator $\mathbf{1}[B_\phi(\xi) = D]$, determined entirely by the look-up table, and the empirical distribution $\mu_{\mathbf{x}}$, determined entirely by the dynamics. The first factor defines the support---which windows permit forced decoupling. The second determines the \emph{occupancy}---the share of the dynamics that visits the support. The framework controls the support; the dynamics control the occupancy.

Two corollaries are immediate. If $n_\text{fd} = 0$, the support is empty: $\theta_\text{fd}(\mathbf{x}) = 0$ for every row of every trajectory---a structural guarantee, not an empirical regularity. If $n_\text{fd} = 16$, the support is $\{0,1\}^4$ itself: $\theta_\text{fd}(\mathbf{x}) = 1$ for every row. Between these extremes, $n_\text{fd}$ sets the size of the region over which dynamics can accumulate forced-decoupled mass.

\subsection{Results}

\begin{table}[tbp]
\centering
\begin{tabular}{lcccc}
\toprule
\textit{Group} & $n$ & $\theta_\text{fd}$ \textit{mean} & $\theta_\text{fd}$ \textit{median} & $\theta_\text{fd} > 5\%$ \\
\midrule
Emergent ($n_\text{fd} \geq 1$, cycle) & 46 & 42.9\% & 39.1\% & 38/46 \\
Constitutive ($n_\text{fd} \geq 1$, no cycle) & 20 & 43.9\% & 30.1\% & 14/20 \\
$n_\text{fd} = 0$ (cycle or no cycle) & 22 & 0.0\% & 0.0\% & 0/22 \\
\bottomrule
\end{tabular}
\caption{\label{tab:dynamics} Dynamical forced-decoupled fraction by structural group (50-seed mean, $N = 101$, $T = 200$, transient $= 50$).}
\end{table}

All 22 rules with $n_\text{fd} = 0$, which include $\phi_{30}$ and $\phi_{110}$, show exactly zero forced-decoupled bonds across approximately $7.6 \times 10^5$ bond observations per rule (50 seeds $\times$ 150 post-transient rows $\times$ 101 bonds per row). This is the $n_\text{fd} = 0$ corollary of Eq.~\eqref{eq:fd_identity}: an empty support guarantees zero occupancy.

\begin{figure}[tbp]
\centering
\includegraphics[width=0.6\columnwidth]{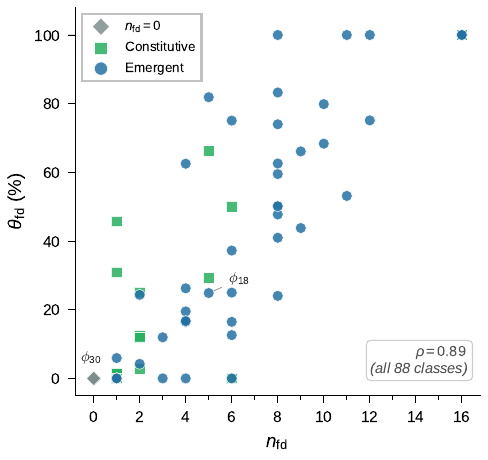}
\caption{\label{fig:scatter} Structural predictor $n_\text{fd}$ versus dynamical $\theta_\text{fd}$ for all 88 equivalence classes (50-seed mean). Emergent rules (cycle present, $n_\text{fd} \geq 1$) in blue; constitutive rules (no cycle, $n_\text{fd} \geq 1$) in green; rules with $n_\text{fd} = 0$ at the origin in grey. Spearman $\rho = 0.89$ ($p < 10^{-31}$).}
\end{figure}

Rules with $n_\text{fd} \geq 1$ show forced-decoupled bonds in proportion to $n_\text{fd}$ (Figure~\ref{fig:scatter}). Nine rules with $n_\text{fd} = 16$ are fully decoupled ($\theta_\text{fd} = 1$)---every window lies in the support. Of the eleven rules with $n_\text{fd} = 1$, nine show $\theta_\text{fd} < 6\%$: the support contains a single window and the dynamics place less than $6\%$ of bonds on that window. The remaining two, both constitutive ($\phi_{57}$, $\phi_{156}$), reach $\theta_\text{fd} > 30\%$, illustrating that simple dynamics can concentrate occupancy on even a minimal support. Between the extremes, $n_\text{fd}$ traces a monotonic gradient. Constitutive and emergent rules follow the same gradient, confirming that $n_\text{fd}$ predicts dynamical decoupling across both regimes.

Across all 88 equivalence classes, $n_\text{fd}$ predicts $\theta_\text{fd}$ with Spearman $\rho = 0.89$ ($p < 10^{-31}$; Table~\ref{tab:correlation}). Within the 46 emergent rules alone, $\rho = 0.83$; within all 57 cycle-positive rules, $\rho = 0.89$. An ablation analysis (Supplementary Information, \S S8) confirmed that $n_\text{fd}$ is the strongest LUT-derived predictor: alternative quantities (PI count, total wildcard count, canalization asymmetry, output weight, canalizing input count) all yield $|\rho| \leq 0.71$.

\begin{table}[tbp]
\centering
\begin{tabular}{lcc}
\toprule
\textit{Scope} & \textit{Pearson $r$} & \textit{Spearman $\rho$} \\
\midrule
Emergent (46 rules) & 0.827 & 0.826 \\
All cycle-positive (57 rules) & 0.870 & 0.886 \\
All 88 equivalence classes & 0.899 & 0.893 \\
\bottomrule
\end{tabular}
\caption{\label{tab:correlation} Correlation between $n_\text{fd}$ and $\theta_\text{fd}$ (50-seed mean). The third row includes the 22 rules with $n_\text{fd} = 0$ at the origin.}
\end{table}

Four rules with $n_\text{fd} \geq 2$ show zero dynamical expression ($\phi_{8}$, $\phi_{128}$, $\phi_{136}$, $\phi_{164}$; all have $n_\text{fd} \leq 6$). The support exists but the dynamics allocate no mass to it---structural permission without dynamical realization. These are not failures of the framework; they mark its boundary, where support size is necessary but not sufficient.

\section{Discussion}
\label{sec:discussion}

\subsection{Forward--backward complementarity}

The epsilon-machine, statistical complexity filters and information-theoretic decomposition identify \emph{what} information-processing structures a CA produces from its dynamics. The tile formalism and the tiling transducer show \emph{how} the rule's Boolean logic distributes causal coupling across the lattice from the look-up table. The two directions are complementary. Rupe and Crutchfield~\cite{rupe2018} note this gap explicitly: generating the equivalence classes over past lightcones from the update rule alone remains an open problem, and ``the only known way to do this is by brute-force simulation and reconstruction''. The tiling transducer provides a partial answer: it derives the spatial coupling architecture---which regions are forced-decoupled, and why---directly from the look-up table, without looking at dynamics.

For $\phi_{18}$, the complementarity is concrete. The epsilon-machine identifies the $(0\Sigma)^*$ domain and embedded particles from spacetime data \cite{hanson1997}. $\mathcal{T}_{\mathrm{eff}}$ identifies the wildcard-emission cycle whose PI-output mapping yields the domain and the zero-wildcard PIs that constitute the particles, computed from the look-up table, without running the dynamics. The two descriptions converge on the same structures through opposite routes.

Langton~\cite{langton1990} proposed that computation in CAs requires three primitive functions: storage, transmission and modification of information. The epsilon-machine's structural elements realize this triad. Domains store information as repeating patterns that persist across time. Particles transmit it as localized signals that propagate through the lattice. Collisions modify it when two signals interact and produce new ones. The tiling framework connects this functional decomposition to the LUT. The wildcard-emission cycle identifies the compositional pathway whose PI-output mapping yields the domain. Forced-coupled PIs generate the localized disruptions that constitute particles. The interaction between the two---where a zero-wildcard PI meets the boundary of a decoupled region---produces the collisions. Langton identified \emph{what} a CA needs to compute; the tiling transducer identifies \emph{how} the LUT generates all three functions through the distribution of causal redundancy across its prime implicants.

\subsection{Relation to overlap-graph constructions}

$\mathcal{T}$ is, at an abstract level, an overlap automaton: it reads symbols left to right, tracking constraints inherited from the previous symbol as finite state. Overlap automata and de~Bruijn graphs have a long history in CA theory~\cite{voorhees1996,sutner1995}. The novelty of $\mathcal{T}$ is not the finite-state overlap architecture itself but three features that distinguish it from the standard construction. First, the input alphabet is not the set of concrete neighbourhoods ($2^3 = 8$ for ECA) but the set of prime implicants---a schema-compressed alphabet whose wildcard positions encode causal redundancy. Second, the state carries resolved overlap demands in a three-valued alphabet $\{0, 1, \#\}$ rather than binary cell values, so that causal freedom propagates through the transducer as an explicit symbol. Third, the pruning from $\mathcal{T}$ to $\mathcal{T}_{\mathrm{eff}}$ removes transitions where a PI is subsumed by a more general one, producing a compressed machine whose structural properties (cycles, emissions, pruning ratio) serve as diagnostics. The de~Bruijn graph of an ECA encodes which neighbourhoods can be adjacent; $\mathcal{T}_{\mathrm{eff}}$ encodes which causal coupling patterns can compose. Riedel and Zenil~\cite{riedel2018} explored Boolean decomposition in a different direction, decomposing ECA rules into prime and composite rules under Boolean composition and identifying minimal generating sets and new universality results. Their decomposition targets computational universality; the tiling transducer targets causal coupling architecture. The two programmes operate on the same rule space but ask different questions.

\subsection{The three-level hierarchy as diagnostic}

Our analysis of the entire ECA landscape reveals three structurally distinct modes of decoupling. Each mode has a different relationship between the look-up table and the dynamics. We find that each produces qualitatively different spacetime structures.

\textit{Constitutive decoupling} requires no spatial mechanism. Each PI individually asserts independence. The result is trivial: constant, identity or XOR-like rules where outputs depend on at most one input. These rules have no coupling to distribute, and their dynamics reflect this---either uniform or purely local.

\textit{Emergent stable decoupling} requires both a spatial mechanism (the wildcard-emission cycle) and temporal closure (self-consistency). Neither suffices alone. The wildcard-emission cycle without self-consistency generates decoupled regions that the update rule immediately destroys. Self-consistency without the cycle limits decoupling to isolated bonds that cannot propagate. The conjunction produces something qualitatively new: self-sustaining domains that persist across time and segment the spacetime into directly decoupled regions. This is the regime of clean domain--particle decomposition. The 46 rules with both a cycle and $n_\text{fd} \geq 1$ (Table~\ref{tab:class}) are candidates for this regime; genuinely self-consistent rules---those whose forced-decoupled triples all survive one application of $\phi$---form a subset (Supplementary Information, \S S9).

\textit{Emergent fragile decoupling} has the spatial mechanism but lacks temporal closure. The 11 rules with a cycle but $n_\text{fd} = 0$ illustrate the extreme case: a wildcard-emission cycle exists, but the support is empty ($S_\phi = \varnothing$), and the support/occupancy identity guarantees exactly zero forced-decoupled bonds. Rules with a cycle and low $n_\text{fd}$ occupy a neighbouring regime: the structural support for decoupling exists. However, the dynamics place less than $6\%$ of bonds on it. The $n_\text{fd}$ gradient from 1 to 16 tracks this transition from negligible to full decoupling.

This hierarchy is a diagnostic tool applicable beyond the specific rules studied here. Given any ECA rule, three computable tests (wildcard-emission cycle, forced-decoupled triples and self-consistency) determine which level the rule occupies and predict the qualitative character of its coupling dynamics. The framework does not explain the full richness of ECA dynamics. It explains one specific feature: whether and how a rule $\phi$ LUT supports persistent directly decoupled regions.

\subsection{From canalization to spatial causal architecture}

The framework developed here extends the canalization programme of Marques-Pita and Rocha~\cite{marquespita2011,marquespita2013} in a specific direction: from single-node redundancy to lattice-wide causal structure. Schema redescription reads which inputs are redundant at each node. The tiling transducer reads what happens when those redundancies compose across adjacent nodes in a spatially homogeneous system.

The tiling framework connects to effective connectivity~\cite{marquespita2013, manicka2022} but operates at a different level. Canalization reduces a network's effective connectivity below its nominal wiring: redundant inputs do not transmit perturbations even when structurally connected. The tiling transducer makes the spatial analogue of this reduction explicit. Forced-decoupling marks bonds where no pair-internal overlap site is an enput for both adjacent transitions under any maximally compressed description. This is a spatial structural analogue of reduced effective connectivity---not yet an intervention-based or information-theoretic equivalence. The wildcard-emission cycle identifies lattice regions where this zero-coupling condition self-sustains across space. Self-consistency identifies where it persists across time.

The three-level hierarchy maps onto this connection. Constitutive decoupling corresponds to rules where canalization is extreme enough that every adjacent pair is forced-decoupled regardless of context. Emergent stable decoupling corresponds to spatial configurations where the wildcard positions of individual PIs align across the tiling to produce self-sustaining regions of forced-decoupled bonds. Emergent fragile decoupling corresponds to configurations where that spatial alignment holds momentarily but the temporal dynamics destroy it.

This bridge between node-level canalization and lattice-level causal architecture suggests a broader programme. In Boolean networks with heterogeneous update functions, each node has its own prime implicant set, and the tiling constraints depend on the specific pair of functions at connected nodes; the structural diagnostics generalize in principle, but the combinatorial landscape grows substantially. The ECA analysis presented here is the homogeneous limit of that more general programme.

\subsection{What the framework does not predict}

The framework characterizes the spatial logic of causal coupling. The structural support $n_\text{fd}$ predicts the dynamical forced-decoupled fraction $\theta_\text{fd}$ with $\rho = 0.89$, and the case studies confirm that the predicted mechanisms match the observed domain structures. What the framework does not predict is Wolfram complexity class: rules from classes II, III and IV intermix across the structural landscape. The support/occupancy identity (\S\ref{sec:identity}) clarifies why. The support is fixed by the LUT; the occupancy depends on how the dynamics distribute mass across windows. This decomposition---separating what a rule \emph{permits} from what the dynamics \emph{realize}---applies wherever a local mechanism defines a set of structurally possible configurations and a global process selects among them. Variation in occupancy---why rules with identical $n_\text{fd}$ produce different $\theta_\text{fd}$---is the open frontier, and may be where the connection to dynamical complexity class resides.

$\mathcal{T}$ reads one row at a time. It captures spatial coupling within a row but not temporal correlations across rows. The period-2 ether of $\phi_{54}$ (a temporal regularity) is invisible to the single-row analysis. An extension to transducer compositions over time steps (which is outside the scope of this paper) would capture temporal correlations at the cost of exponential state-space growth.

The current analysis is restricted to elementary (radius-1, binary) CAs. The prime implicant decomposition and overlap resolution generalize to higher radius and larger alphabets. The state space of $\mathcal{T}$ grows as $|\{0, 1, \#\}|^{2r}$---manageable for radius 2 ($3^4 = 81$ states) but increasingly costly beyond that.

\subsection{Limitations}

The dynamical validation averages over 50 random seeds (seeds 42--91) for all 88 equivalence classes. Individual-rule $\theta_\text{fd}$ values show low variance (typical standard deviation 1--5 percentage points; see Supplementary Information, Table~S10), confirming that the reported correlations are not seed-dependent artefacts.

The framework treats all PIs of equal wildcard count as equally valid descriptions. Whether the dynamics preferentially realize some PIs over others (through global constraints not captured in the four-cell window) remains an open question. The four rules with zero dynamical expression despite $n_\text{fd} \geq 2$ (\S\ref{sec:dynamical-validation}) demonstrate that structural permission does not guarantee dynamical realization even when the support contains up to six windows.

\section{Conclusion}

The tile formalism, the Tiling Transducer and the three-way bond classification provide a forward model for reading the causal architecture of cellular automata from the look-up table. Dynamical validation across all 88 equivalence classes confirms the framework's predictions: the structural count of forced-decoupled windows $n_\text{fd}$ (the support size of the bond-classification map $B_\phi$) predicts the dynamical forced-decoupled fraction $\theta_\text{fd}$ with Spearman $\rho = 0.89$ ($p < 10^{-31}$; 50-seed mean). For the 22 rules with $n_\text{fd} = 0$, the support/occupancy identity guarantees zero forced-decoupled bonds as a theorem; for rules with $n_\text{fd} = 16$, it guarantees full decoupling. Between these extremes, $n_\text{fd}$ traces a continuous gradient from negligible to full decoupling---a quantitative prediction from the look-up table alone.

The framework connects the look-up table to the dynamics it produces through a single mechanism. The distribution of causal redundancy across the prime implicants determines the tile catalogue. The catalogue determines $\mathcal{T}$. The transducer's structural properties (its cycles, its emissions, its pruning) predict the spatiotemporal coupling patterns. The look-up table encodes not just what the rule computes but \emph{how} it distributes causal coupling across space, and that distribution determines whether the rule's dynamics decompose into persistent directly decoupled regions.

Schema redescription~\cite{marquespita2011,marquespita2013} reads the causal structure of individual nodes. Effective connectivity~\cite{manicka2022} links that structure to network-level dynamical regimes. The Tiling Transducer fills the gap between the two: it reads how node-level redundancy composes into lattice-wide causal architecture. The ECA landscape analysed here is the simplest case---homogeneous, one-dimensional, binary. The logic of tile composition, bond classification and cycle analysis is not restricted to this case. Extending the framework to higher-dimensional lattices, larger alphabets and heterogeneous Boolean networks is a natural next step; relating direct coupling weight to explicit intervention-based causal measures and partial-information-decomposition quantities is another. The present paper establishes the structural foundation for that programme: a forward pipeline from local logical redundancy to a predictive theory of emergent mesoscopic organization.

\bibliographystyle{unsrtnat}
\bibliography{references}

\end{document}